# Multipath Effects on Frequency-Locked Loops (FLLs) and FLL-derived Doppler Measurements

Liangchun Xu, *Tufts University*
Jason Rife, *Tufts University*

## ABSTRACT

This paper investigates the impact of non-line-of-sight (NLOS) and multipath signals on frequency-locked loops (FLLs), which are commonly used to obtain Doppler shift measurements for velocity estimation in radio navigation. A salient result is that, in the absence of a direct signal, a single NLOS signal does not corrupt Doppler observables if the angle-of-arrival (AOA) is known. Another striking result is that, in a multipath environment involving two signals, the arctangent discriminator (averaged over the beat frequency) tracks only the higher amplitude signal. Our investigation has particular significance for radio navigation in urban environments, either using global navigation satellite system (GNSS) signals or ground-based cellular signals.

## INTRODUCTION

Non-line-of-sight (NLOS) and multipath signals are common nuisances for wireless communication and positioning systems [1, 2]. A NLOS signal is an indirect signal arriving alone at the receiver, while multipath consists of two or more signal rays arriving together at a receiver from the same source [3]. Multipath always includes at least one reflected (or indirect) signal and typically also includes the direct line-of-sight (LOS) signal. From the perspective of radionavigation, NLOS signals and multipath signals are usually considered detrimental to both the pseudorange and the Doppler measurements. As such, various fault detection and exclusion (FDE) methods have been developed to identify and discard NLOS and multipath signals [1, 2, 4].

Although multipath effects are well studied for code tracking in delay-locked loops (DLLs) and for carrier-phase tracking in phase-locked loops (PLLs) [5, 6, 7, 8, 9], multipath effects on carrier-frequency tracking in frequency-locked loops (FLLs) are not fully understood. In this paper, we investigate the effects of multipath on FLLs. We seek to identify the mechanisms that degrade FLL-derived Doppler observables under multipath conditions. A key result is that the Doppler measurement for an individual NLOS signal can be interpreted geometrically using only angle of arrival (AOA), without knowledge of reflector geometry, so long as reflectors are nearby and stationary. For multipath signals consisting of two components, the amplitude ratio between the two arriving signal components plays a pivotal role in FLL performance, as we first investigated in [10]. As we saw [10] and as we expand on in this paper, the FLL tracks the Doppler shift of the signal component with larger amplitude. This generalization is true at least to the extent that the loop filter suppresses periodic perturbations at the beat frequency, which result from the Doppler-shift difference between the two arriving signal components. This generalization is striking because it indicates the FLL essentially tracks only one component of two-ray multipath: either the LOS signal or an NLOS signal component, but not both.

Better understanding the impact of multipath on FLLs has several practical implications. Importantly, error models enable better characterization of the degradation of FLL-derived Doppler measurements, particularly in urban applications where multipath is strong [11, 12, 13, 14]. Further, multipath detectors using Doppler shift measurements can be designed to exclude detrimental multipath measurements [15]. In the longer term, a better understanding of multipath on Doppler observables might also help in exploiting NLOS signals as additional signals-of-opportunity to improve navigation performance. For example, leveraging the observation that Doppler measurements are not corrupted by a reflection from a stationary surface, a multi-input, multi-output (MIMO) antenna-based system with beamforming capabilities could in concept exploit NLOS measurements as signals of opportunity [16].

Our investigation of multipath effects on FLLs is structured as follows. First, as background, we review a general model of multipath. Second, we use theory and numerical models to characterize the effect of two-ray multipath on an FLL. We present our results in terms of averaging over the beat period between the two rays, and subsequently in more nuanced terms, in terms of temporal variations within the beat period. In the following, we provide context for our theoretical models and to demonstrate the deleterious impact of multipath on the Doppler observable for a conventional receiver. A brief discussion and summary conclude the paper.

## BACKGROUND

This section reviews standard approaches for modeling multipath in radionavigation applications. As a starting point, let us consider a case like that illustrated in Fig. 1, where a transmitted signal can take both a direct LOS path to the receiver and an indirect NLOS path. We assume, the NLOS signal reflects from a nearby, stationary surface.

Both signal pathways are generated from the same transmitted signal. The transmitted signal $s_T$ consists of a pseudorandom code $c$ modulated on a carrier wave that oscillates at a frequency $\omega$:

$$s_T(t) = A_T e^{j(\omega t)} c(t) \tag{1}$$

Here $A_T$ is the amplitude of the transmitted signal and $t$ is time. The components of the compound signal that arrives at the user antenna are labeled $s_l$, where the index $l$ is 0 for the LOS path and a positive integer for each NLOS path. Each path may have a distinct Doppler shift $\omega_{d,l}$. For instance, the direct signal component features a Doppler shift $\omega_{d,0}$, which depends on the relative motion between the receiver and the transmitter. Note that in the signal model we do not consider the navigation bits because we assume that the discriminator is not computed across bit transitions [6].

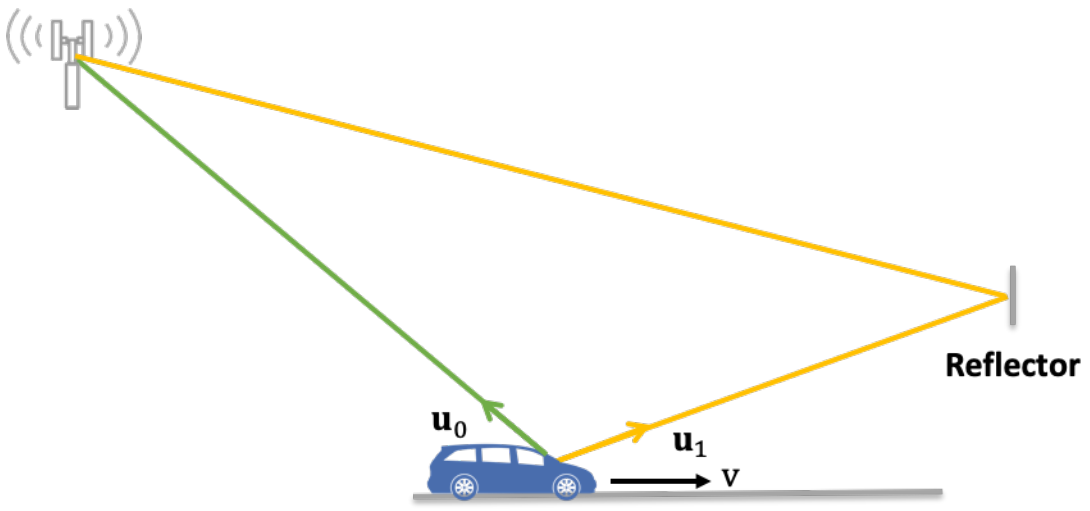

Fig. 1. Example of a multipath signal consisting of a direct LOS signal (green) superposed with an indirect NLOS signal (yellow). Outward unit vectors describe bearing from the receiver antenna to the source of the transmission in the case of the LOS signal ($\mathbf{u}_0$) or to the reflector in the case of the NLOS signal ($\mathbf{u}_1$). The velocity of the car is denoted by $\mathbf{v}$

A model for the LOS signal component $s_0$ may be written in the following form.

$$s_0(t) = A_0(t) e^{j\left((\omega+\omega_{d,0})t+\phi_0\right)} c(t-\tau_0) + n(t) \tag{2}$$

Here the Doppler shift $\omega_{d,0}$, carrier phase offset $\phi_0$ and the code time delay $\tau_0$ depend on time-of-flight and possibly other propagation effects, such as phase shifts due to reflection. The direct component is degraded by spreading losses, which determine its amplitude $A_0$, and by thermal noise, which is modeled as an additive random variable $n$.

The direct signal component combines with indirect NLOS signals arriving at the user antenna. In Fig. 1, there is only one NLOS signal, but in general there may be a number $L$ of indirect signals that arrive at the user antenna. Modeling the NLOS components in a manner analogous to (2) and summing with the LOS signal gives the total received signal $s$:

$$s(t) = \sum_{l=0}^{L} \left( A_l(t) e^{j\left((\omega+\omega_{d,l})t+\phi_l\right)} c(t-\tau_l) \right) + n(t) \tag{3}$$

Note that the random noise variable $n$ is modeled as a single term for the user receiver, and not as separate terms for each signal component. By analyzing (3), we can infer the effects of multipath on receiver observables, such as pseudorange and Doppler measurements, as a function of the parameters for each raypath ($A_l, \phi_l, \omega_{d,l}, \tau_l$).

In the following sections, we will analyze the effect of multipath on FLL-derived Doppler estimation for the two-ray case; first, as context, it is instructive to review existing models describing how multipath degrades the pseudorange observable.

Models of the impact of multipath on the pseudorange observable have been studied extensively for GPS applications [17], where the pseudorange observable is most typically obtained as the output of a DLL. Multipath distorts the correlation peak tracked by the DLL and, most typically, increases the apparent pseudorange as measured by the DLL (up to 1.5 code chips delay). Some multipath error for code pseudorange can reach tens of meters.

The code multipath effect can be visualized as a distortion of the ideal correlator peak for the direct signal, arriving at time $\tau_0$ as described in (2), due to the arrival of NLOS signals, each delayed by a time $\tau_l$ as described in (3). This effect is shown below in Fig. 2, which plots correlator outputs as a function of the chip offset, measured in chips (where one chip corresponds to one bit of code). For a pseudorandom noise code that consists of a series of random digital bits, the individual autocorrelation functions are predominantly triangular in shape, as shown by the direct (dashed) and NLOS (solid) signals, which are considered individually on the left side of Fig. 2. When the LOS and NLOS signals are superposed, however, the combined correlator output is no longer triangular, as shown on the right side of Fig. 2, which shows the output of the correlator to the superposed LOS and NLOS signals, with the peak value normalized to one. The correlation peak is shifted later in time [3], nearly a full chip later in this case. When the peak location is extracted, for instance using an early-late discriminator as an input to a DLL [5], then the apparent time-of-flight is extended and the observed pseudorange is artificially increased.

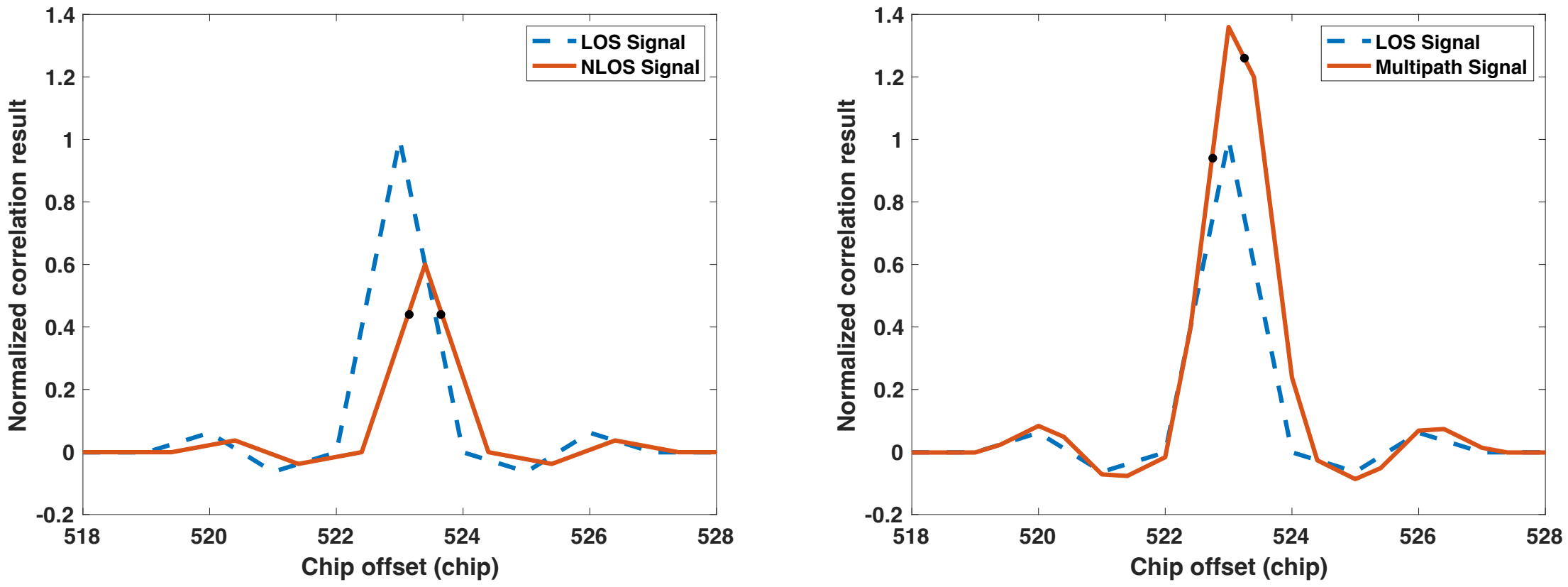

Fig. 2. An example of auto-correlation for the NLOS signal (left) and the multipath signal (right). The peak of the received composite signal is shifted significantly in the case of NLOS versus multipath [3]. The figure shows the special case when the LOS signal is stronger than the NLOS signal

Significant research has also been directed to the study of multipath effects on carrier-phase pseudorange measurements as acquired by a PLL [5, 6, 7, 8, 9]. For two-ray multipath where the LOS component has greater power than the NLOS component, it is well known that the carrier-phase error is no more than a quarter cycle (less than 5 cm) [5]. However, this assumption may be violated, particularly in urban environments where buildings may occlude the LOS measurement but not the NLOS measurement, and so it is important to recognize that carrier-phase multipath errors may significantly exceed 5 cm in the general case. Moreover, carrier-phase tracking is fragile under severe multipath, which makes it difficult to ensure the availability of the carrier-phase measurements and even more difficult to maintain an integer ambiguity fix.

In contrast, the multipath effect on frequency tracking is not investigated yet.

For applications requiring reliable velocity estimation, the FLL-derived Doppler shift observable has several advantages as a complement to PLL and DLL-derived observables. The advantages are particularly salient for urban environments with strong multipath. First, an FLL is more resilient than a PLL in the face of low carrier-to-noise ratio ($C/N_0$) induced by structures, because the FLL uses incoherent processing in contrast with the PLL which uses coherent processing [18]. In this environment, the FLL can tolerate larger tracking errors and, unlike the PLL, is not subject to cycle-slip induced errors. Second, Doppler observables are much less noisy than velocity estimates obtained by differencing DLL-derived pseudoranges, by an order of magnitude or more. In order to leverage these benefits, we introduce a mathematical model for multipath effects on FLLs in the following section. For the most part, prior work on FLLs focuses on low signal-to-noise ratio conditions [19, 20], and specifically on events when only one NLOS signal is tracked. In this paper, we consider more general cases, such as multipath cases where both LOS and NLOS signals are present simultaneously. Previous researchers have also focused more on hardware studies analyzing the correlator outputs [19, 21-23], by contrast we take a theoretical approach to analyze the whole tracking loop, which is the key enabler for us to analyze tracking of events where LOS and NLOS inputs are superposed.

## MULTIPATH EFFECTS ON FREQUENCY-LOCKED LOOPS

This section provides, for the first time, a theoretical characterization of how multipath affects the FLL output and its Doppler observable. Acknowledging that many approaches exist for implementing an FLL, in this work, we attempt to generalize the FLL somewhat, to capture a wide range of FLL implementations. Our generalized model consists of sequence of four signal processing steps including carrier and code wipeoff, integration and dump, discriminator, noncoherent moving average, and loop filter blocks. These blocks are shown in Fig. 3 The output of the loop filter block is used to update the estimated Doppler frequency, which is in turn used to steer the numerically controlled oscillator (NCO) and adjust the code replica that is compared against the received signal as an input to coherent averaging.

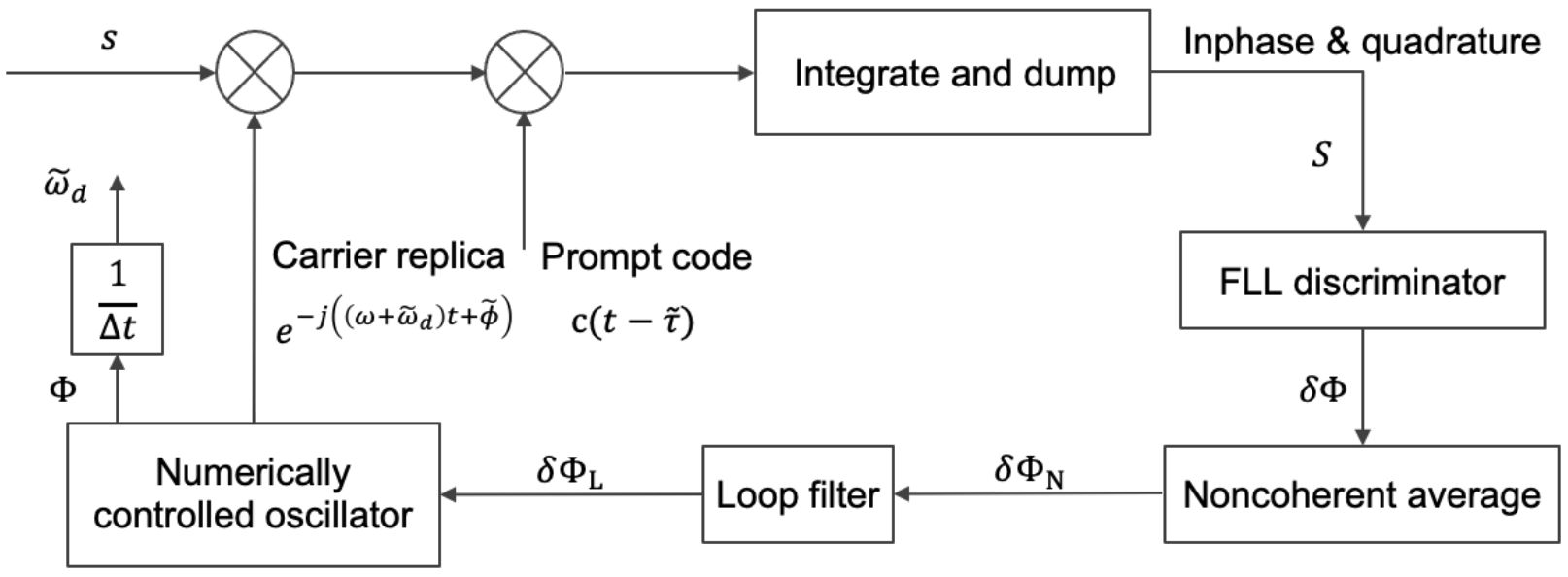

Fig. 3. Block diagram of a typical FLL

**Correlation Process**

As shown in Fig. 3, the first three FLL processing steps are carrier wipeoff, code wipeoff, and integration. Carrier and code wipeoff removes the code modulation $c(t)$ and the nominal carrier and Doppler frequencies from the input signal (3). This leaves the frequency-tracking error $\delta\omega_l$, which the FLL loop attempts to zero. If the frequency tracking error is small (as is typical), the superposed sinusoids for all ray paths pass unmodified through integration. The resulting output $S$ of the whole correlation process is:

$$S(t) = \sum_{l=0}^{L} \left(B_l(t) e^{j((\delta\omega_l)t+\phi_l)}\right) + n(t) \tag{4}$$

For completeness Appendix 1 derives this pass-through result for the correlation process, which relates the postcorrelation signal (4) to the precorrelation signal (3). Note the amplitude is modified slightly (labeled $B_l$ instead of $A_l$) to account for code wipeoff. Since quantifying noise is not a major focus of this paper, we simply reuse the term $n$ to model the additive noise.

**FLL Observable for LOS-only Signal**

Next consider the discriminator block, which takes as input the signal $S$ given by (4). Although (4) models an arbitrary number $L$ of NLOS signals, it is useful to develop baseline equations for the discriminator by considering the simplest case when $L = 0$, meaning that only the LOS component is present.

The discriminator processes $S$ to obtain a frequency error $\delta\omega_0$. The frequency error is the phase angle of the product of two signals: the post-correlation sample $S(t)$ and its complex conjugate at an earlier time, labeled $S'(t - \Delta t)$, divided by $\Delta t$.

$$\delta\omega_0 = \frac{\delta\Phi}{\Delta t} = \frac{\angle S(t)S'(t - \Delta t)}{\Delta t} \tag{5}$$

Here $\Delta t$ is a small offset that is selected as an FLL design parameter. For instance, if $\Delta t$ is set to the integration time, then $S(t - \Delta t)$ is simply the last sampled postcorrelation signal. Since $\Delta t$ is constant, $\delta\Phi$ is analyzed in the following text.

Equation (5) is exact when an arctangent discriminator design is employed as shown in Appendix 2. Other common FLL discriminator designs, approximate (5) in a linearized sense. For instance, the cross-product discriminator (as described in [5]) is essentially equal to $\delta\Phi$ when the phase angle is small. To keep our analysis as simple as possible, we will adopt the assumption of an arctangent discriminator, where (5) is exact.

To extract the phase angle from a complex number, it is common to use the arctangent function defined over the range from $-\pi$ to $\pi$:

$$\delta\Phi = \operatorname{atan2}\left(\frac{\operatorname{Im}\big(S(t)S'(t-\Delta t)\big)}{\operatorname{Re}\big(S(t)S'(t-\Delta t)\big)}\right) \tag{6}$$

To analyze the phase angle for the case of an LOS signal only, we can expand (6) using (4), with $L$ set to 0. Assuming the Doppler error changes little over the interval $\Delta t$ (usually 1-5 *ms* for strong GPS L1 signals), then the phase angle $\phi_0$ cancels leaving only:

$$\delta\Phi = (\delta\omega_0)\Delta t \tag{7}$$

In the FLL loop, as shown in Fig. 3, each difference $\delta\Phi$ measured by the discriminator reflects the instantaneous Doppler error between the signals of current and previous epochs. These instantaneous errors are passed through the filtering blocks (labeled "noncoherent average" and "loop filter" in the graphic, with outputs $\delta\Phi_N$ and $\delta\Phi_L$ respectively) and accumulated at the NCO to estimate Doppler as $\Phi$.

Assuming linear filtering, the FLL-observable can be computed equivalently by switching the order of the filtering and accumulation operations. As such, we will define an accumulated discriminator variable $\Phi$, which represents integration by the NCO if accumulation were implemented before filtering. The estimate $\Phi$ converges (at least in the absence of noise) when the tracking error $\delta\omega_0$ becomes zero, which in turn implies that the frequency of the local carrier replica is $\omega_R = \omega + \omega_d$. Assuming the integral $\Phi$ is initialized correctly, the converged value of the accumulated discriminator is

$$\Phi = \int (\delta\Phi)\mathrm{dt} = (\omega + \omega_d)\Delta t \tag{8}$$

Here the central frequency $\omega$ is a known offset that is included in initialization of the loop filter. Absent rapid changes in the Doppler frequency, the filtering blocks in Fig. 3 will not have any significant effect on the values $\delta\Phi$ (that is $\delta\Phi \approx \delta\Phi_N \approx \delta\Phi_L$), so the estimate provided by (8) is a good approximation of the estimate accumulated by the NCO. Manipulating (8) we obtain a Doppler frequency estimate as:

$$\tilde{\omega}_d = \frac{\Phi}{\Delta t} - \omega \tag{9}$$

Here the tilde superscript implies an estimated quantity. The estimated Doppler frequency reflects the combined contributions of the receiver and transmitter motion. As a reasonable approximation, the Doppler contributions of the receiver and transmitter, $\omega_r$ and $\omega_t$, sum to give a combined Doppler shift for the direct signal:

$$\omega_d \approx \omega_r + \omega_t \tag{10}$$

Combining (9) with (10), and assuming the transmitter motion contribution is well characterized, we can estimate the receiver contribution to the Doppler shift as $\tilde{\omega}_r$:

$$\tilde{\omega}_r = \frac{\Phi}{\Delta t} - \omega - \omega_t \tag{11}$$

The transmitter contribution $\omega_t$ is assumed to be well characterized in the sense that it can be computed from broadcast parameters (in the case of GNSS) or that it is zero (in the case of a stationary terrestrial base station). For satellite transmitters, we additionally assume that the reflector is near the receiver (within a few hundred meters) such that the propagation direction from the transmitter to the receiver is essentially parallel to that from the transmitter to the reflection point. This avoids differential satellite-induced Doppler that can occur when the reflector is far from the receiver [20]. Note in this case we consider only large stationary reflectors, and so (10) does not need to be modified to account for reflector motion.

The receiver's Doppler contribution can be used to infer the user-receiver velocity. The receiver Doppler contribution $\omega_r$ is related to the projection of the receiver velocity $\mathbf{v_r}$ onto the unit LOS vector to the receiver:

$$\omega_r = -\frac{2\pi}{\lambda}\mathbf{v_r} \cdot \hat{\mathbf{u}}_r = -\frac{2\pi\|\mathbf{v}_r\|}{\lambda}\cos(\theta_r) \tag{12}$$

Here the carrier wavelength is $\lambda$ and unit pointing vector in the direction of the arriving signal is $\hat{\mathbf{u}}_r$. Two forms of the velocity relationship are presented; the second replaces the dot product with an arrival angle $\theta$ between the unit vector from receiver to satellite and the velocity vector.

**Period-Averaged FLL Observable for NLOS and Multipath Cases**

This subsection extends the previous FLL discriminator to include cases other than the LOS-only case. Specifically, we will now consider NLOS and two-ray case where, nominally, one of the rays is the LOS signal and the other, the NLOS signal. These multipath cases can be modeled using the first two terms of (4).

In order to evaluate two-ray multipath effects on the FLL, we must consider how the postcorrelation signal (4) maps through the discriminator and into the accumulated phase $\Phi$, as given by (8). This section builds from the LOS case of last section toward the two-ray multipath case by considering several intermediate cases: the case of only the NLOS signal with no LOS signal, the case of two-rays (LOS and NLOS) and a stationary receiver, and finally the two-ray case for a moving receiver. Each of these three cases is summarized below by a remark. The remarks are intended to provide intermediate results to support future experimental verification of the model through data collection in urban environments.

*Remark 1: [NLOS case] When one NLOS signal arrives at a receiver after reflecting from a stationary surface and when the LOS signal is absent, the Doppler observable is proportional to receiver velocity projected on the NLOS path.*

*Discussion:* As shown in Fig. 1, the NLOS signal arrives at the receiver along the raypath from the receiver to the location of the reflection. If the unit vector along the arriving raypath is labeled $\hat{\mathbf{u}}_{r,l}$ then (12) can be generalized to describe the projection of the vehicle velocity onto that raypath direction.

$$\omega_{r,l} = -\frac{2\pi}{\lambda}\mathbf{v}_{\mathbf{r}} \cdot \hat{\mathbf{u}}_{r,l} = -\frac{2\pi\|\mathbf{v}_r\|}{\lambda}\cos\left(\theta_{r,l}\right) \tag{13}$$

Each raypath has a different Doppler shift $\omega_{r,l}$ due to receiver motion, because each raypath has a distinct angle $\theta_{r,l}$ relative to the direction of motion of the receiver.

If only one NLOS signal is received ($l = 1$) and no direct signal is received, then it is easy to substitute (13) for (12) in the analysis of the prior section. By extension, the FLL will estimate $\omega_{r,1}$ the Doppler frequency obtained by projecting the receiver velocity $\mathbf{v}_{\mathbf{r}}$ on to the NLOS raypath unit vector $\hat{\mathbf{u}}_{r,1}$.

*Remark 2: [Zero-speed multipath case] When the receiver and reflective surfaces are stationary, multipath signals have no effect on the Doppler observable.*

*Discussion:* When the receiver is stationary, the receiver speed is $\|\mathbf{v}_r\| = 0$. The receiver-motion contribution to Doppler shift $\omega_{r,l}$ can be computed by (13) for all raypaths to be

$$\omega_{r,l} = 0 \tag{14}$$

The total Doppler consists of this receiver-motion contribution $\omega_r$ as well as a transmitter-motion contribution $\omega_t$. In practice, the transmitter-motion contributions are either known in the case of GNSS applications, or the transmitter is simply not moving (no Doppler contribution) in the case of stationary terrestrial antennae, as in a cellular phone network. Thus, when the receiver and reflective surfaces are motionless, the total Doppler shift is $\omega_d \approx \omega_t$. This is again a trivial extension of the analysis in the prior section. When the Doppler shift is equal for all raypaths, the time-varying exponential is the same in all terms of (4) and can be factored out of the summation as in:

$$S(t) = e^{j(\delta\omega)t}\sum_{l=0}^{L}\left(B_l(t)e^{j\phi_l}\right) \tag{15}$$

This is a summation of sine waves with the same frequency but different phase shifts. Equation (15) can be simplified by writing

$$S = B_{sum}e^{j(\delta\omega)t} \tag{16}$$

with

$$B_{sum} = \sum_{l=0}^{L} B_l e^{j\phi_l} \tag{17}$$

The phase of (16) can be accumulated to provide a $\Phi$ value through (8) and subsequently a projected-velocity estimate through (12). In short, when the receiver is stationary, multipath in no way biases or distorts the FLL-based velocity observable.

It is important to note that Xie and Petovello [18] arrived at a similar result; however, their analysis applied to the single ray cases (NLOS or LOS only). Here we've generalized the analysis to apply to multipath signals composed of two or more rays.

*Remark 3: [Nonzero-speed multipath case]. When multipath arrives at a moving receiver via two raypaths and when reflective surfaces are stationary, the period-average of the arctan discriminator tracks the signal component with higher power.*

*Discussion:* In this case we consider two-ray multipath but with a moving receiver. For this case the input to the discriminator is again modeled by the first two terms of (4). These terms can be written as

$$S = B_0 e^{j\theta_0} + B_1 e^{j\theta_1} \tag{18}$$

where

$$\theta_l = (\delta\omega_l)t + \phi_l, \qquad l = 0,1 \tag{19}$$

and, according to Appendix 1, each delta frequency $\delta\omega_l$ is a difference of the replica frequency $\omega_R$ from the received frequency for a given component

$$\delta\omega_l = \omega_l - \omega_R, \qquad l = 0,1 \tag{20}$$

In order to analyze how the arctangent discriminator processes (18), it is useful to factor the signal by introducing two variables of substitution: $\gamma$, which describes a phase difference, and $\mu$, a phase average.

$$\gamma = \frac{\theta_0 - \theta_1}{2}, \qquad \mu = \frac{\theta_0 + \theta_1}{2} \tag{21}$$

Substituting these variables into (18) and using algebraic identities gives

$$S = e^{j\mu}(B_0 e^{j\gamma} + B_1 e^{-j\gamma}) \tag{22}$$

As noted above, the discriminator compares the prompt signal (21) to a delayed version of itself, where the delay is $\Delta t$. As such, there is an analogy to (21) for the delayed signal, where $S(t - \Delta t) = e^{j\mu_\Delta}(B_0 e^{j\gamma_\Delta} + B_1 e^{-j\gamma_\Delta})$ with $\gamma_\Delta$ describing the phase difference and $\mu_\Delta$ the phase average for the delayed signal. Using these definitions, the product $S(t)S'(t - \Delta t)$ becomes

$$S(t)S'(t - \Delta t) = e^{j(\Delta\mu)}\big((B_0^2 + B_1^2)\cos(\Delta\gamma) + 2B_0B_1\cos(\delta) + j(B_0^2 - B_1^2)\sin(\Delta\gamma)\big) \tag{23}$$

Three intermediate parameters are used in this expression: the difference $\Delta\mu$ between the prompt and delayed phase averages, the difference $\Delta\gamma$ between the prompt and delayed phase differences, and the sum $\delta$ for the prompt and delayed differences.

$$\Delta\mu = \mu - \mu_\Delta \tag{24}$$
$$\Delta\gamma = \gamma - \gamma_\Delta \tag{25}$$
$$\delta = \gamma + \gamma_\Delta \tag{26}$$

These parameters can be expanded using (19) and (21) to show they are equivalent to

$$\Delta\mu = \frac{\delta\omega_1 + \delta\omega_2}{2}\Delta t \tag{27}$$
$$\Delta\gamma = \frac{\delta\omega_0 - \delta\omega_1}{2}\Delta t \tag{28}$$

$$\delta = (\delta\omega_0 - \delta\omega_1)t - \Delta\gamma + \phi_0 - \phi_1 \tag{29}$$

Using the definition of $\delta\omega_0$ and $\delta\omega_1$ in (20), we can rewrite $\Delta\gamma$ and $\delta$ in terms of absolute frequencies rather than tracking errors, which gives

$$\Delta\gamma = \frac{\omega_0 - \omega_1}{2}\Delta t \tag{30}$$

$$\delta = (\omega_0 - \omega_1)t - \Delta\gamma + \phi_0 - \phi_1 \tag{31}$$

The key variable here is $\delta$, which is an apparent contribution to accumulated Doppler due to the interaction of the two signals received along different raypaths. The discriminator output is the phase angle of (23), which can be rewritten as

$$S(t)S'(t-\Delta t) = e^{j\left(\Delta\mu + \operatorname{atan2}\left(\frac{(B_0^2 - B_1^2)\sin(\Delta\gamma)}{(B_0^2 + B_1^2)\cos(\Delta\gamma) + 2B_0B_1\cos(\delta)}\right)\right)} \tag{32}$$

The arctangent discriminator (6) extracts the phase angle of (32), so the discriminator output $\delta\Phi$ can be written

$$\delta\Phi = \Delta\mu + \operatorname{atan2}\left(\frac{(B_0^2 - B_1^2)\sin(\Delta\gamma)}{(B_0^2 + B_1^2)\cos(\Delta\gamma) + 2B_0B_1\cos(\delta)}\right) \tag{33}$$

The only time-dependent term in (33) is $\cos(\delta)$. More specifically, we note that $\delta$ is a linear function of time, according to (31), and that the arctangent operation makes the overall time variations periodic. The periodicity $T_{beat}$ occurs with a beat frequency set by (31), where the term *beat frequency* refers to the difference of the frequencies on the two ray paths [24]. Specifically, the beat frequency and the beat period are defined as

$$\omega_{beat} = |\omega_0 - \omega_1| \tag{34}$$

$$T_{beat} = \frac{2\pi}{\omega_{beat}} \tag{35}$$

In contrast with the single ray analyses above, where the discriminator output $\delta\Phi$ converged to zero in steady-state (with the noise-free tracking error going to zero), $\delta\Phi$ never goes to zero in the two-ray case. Instead, the tracking loop converges to a periodic steady-state, with the input signal exciting the tracking loop at the beat frequency. In order to analyze this periodic steady-state behavior, we break the tracking error into two parts, a DC component and a deviation from DC.

The remainder of this section will focus on estimating the DC component of the accumulated phase $\Phi$. The DC component can be obtained by calculating the time average phase change, denoted with brackets as $\langle\delta\Phi\rangle$.

$$\langle\delta\Phi\rangle = \frac{1}{T_{beat}}\int_0^{T_{beat}} \delta\Phi \, dt = \Delta\mu + \frac{1}{2\pi}\int_0^{2\pi} \operatorname{atan2}\left(\frac{(1-\beta^2)\sin(\Delta\gamma)}{(1+\beta^2)\cos(\Delta\gamma) + 2\beta\cos(\delta)}\right) d\delta \tag{36}$$

In this formula, the amplitude ratio $\beta = B_1/B_0$ has been introduced to combine the two amplitudes $B_l$ and reduce the number of parameters by one. Because $\delta$ increases linearly with time, we have replaced the time-average over the period with an average of the angle $\delta$ over its full range of values (from 0 to $2\pi$).

The integral in (36) does not have a known analytical solution; however, we have been able to use numerics to obtain a solution. Specifically, we computed a normalized integral $f(\beta, \Delta\gamma)$ defined as

$$f(\beta, \Delta\gamma) = \frac{1}{2\pi(\Delta\gamma)}\int_0^{2\pi} \operatorname{atan2}\left(\frac{(1-\beta^2)\sin(\Delta\gamma)}{(1+\beta^2)\cos(\Delta\gamma) + 2\beta\cos(\delta)}\right) d\delta \tag{37}$$

Remarkably, numerical integral produced, to machine precision, the following result:

$$f(\beta, \Delta\gamma) = \begin{cases} -1 & if\ \beta > 1 \\ 0 & if\ \beta = 1 \\ 1 & if\ \beta < 1 \end{cases} \tag{38}$$

Substituting this result into (36) gives a formula for the DC component of the accumulated phase:

$$\langle \delta\Phi \rangle = \begin{cases} (\delta\omega_1)\Delta t & if\ \beta > 1 \\ \dfrac{\delta\omega_1 + \delta\omega_2}{2}\Delta t & if\ \beta = 1 \\ (\delta\omega_0)\Delta t & if\ \beta < 1 \end{cases} \tag{39}$$

According to (8), then the accumulated phase is equal to

$$\langle \Phi \rangle = \int \langle \delta\Phi \rangle\, \mathrm{dt} = \begin{cases} (\omega + \omega_{d,1})\Delta t & if\ \beta > 1 \\ \left(\omega + \dfrac{\omega_{d,0} + \omega_{d,1}}{2}\right)\Delta t & if\ \beta = 1 \\ (\omega + \omega_{d,0})\Delta t & if\ \beta < 1 \end{cases} \tag{40}$$

Removing the central frequency by substituting (40) into (9), we find that the period averaged Doppler estimate is

$$\langle \widetilde{\omega}_d \rangle = \begin{cases} \omega_{d,1}, & if\ \beta > 1 \\ \dfrac{1}{2}(\omega_{d,0} + \omega_{d,1}) & if\ \beta = 1 \\ \omega_{d,0} & if\ \beta < 1 \end{cases} \tag{41}$$

In other words, the DC component of FLL tracking matches the NLOS Doppler when the power of the NLOS component is larger ($\beta > 1$) and the LOS Doppler when the power of the LOS component is larger ($\beta < 1$). The intermediate case is a special limiting case, where the step function (38) is singular, and where the value must be approximated as the average of the function on either side of the singularity. In short, in the two-ray case absent noise, the DC component of FLL tracking extracts the Doppler of the stronger signal component.

Note that our numerical approach to evaluating $f(\beta, \Delta\gamma)$ was to compute its value over a range of discrete $\beta$ from $10^{-4}$ to $10^4$ on a log scale and over a range of discrete $\Delta\gamma$ from infinitely small ($\sim 10^{-6}$) to its theoretical limit $\pi/2$. Over 3000 cases were computed, and all agreed with (38) to numerical precision. In all case, the dummy variable $\delta$ was integrated from 0 to $2\pi$. The final integral results are shown in Fig. 4.

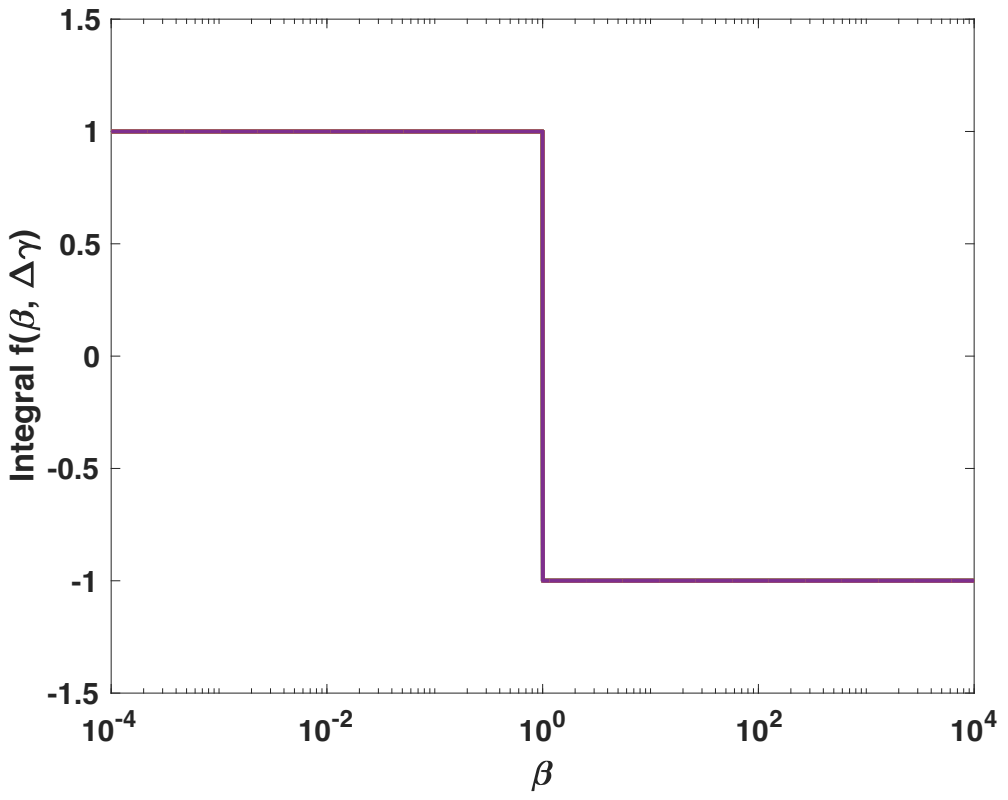

Fig. 4. Integral $f(\beta, \Delta\gamma)$ for all $\beta, \Delta\gamma$ considered as a numerical proof of the step function in (38)

The plot shows a clear step function with respect to $\beta$. Interestingly, there is no variation in $f(\beta, \Delta\gamma)$ for changes in $\Delta\gamma$. The curves for different $\Delta\gamma$ are superposed in the figure, but no distinction is evident since all of the cases produce precisely the same curve.

Although receiver noise is not considered in this paper, our preliminary results indicate that under the primary effect of noise is to smooth the corners of the step in Fig. 4, so that it takes on a reversed S shape with its inflection point at $\beta = 1$. A more detailed noise analysis will be left to a future paper.

## PERIODIC DEVIATIONS RELATIVE TO DC VALUE

In the case when a receiver is moving at a constant speed and experiences two-ray multipath, the FLL-derived Doppler observable varies periodically in time, as observed in the prior section. This section characterizes these oscillations relative to the DC component of the FLL output. To study these periodic deviations, let us return to the model of the arctangent discriminator, as characterized by (33).

For a receiver moving at a constant velocity, the periodic deviations of the arctangent discriminator output are driven by the cosine term containing the variable $\delta$, which is the only time-dependent variable in (33). The variable $\delta$ is a linear function of time *t*, as characterized by (29), so the value of the cosine term is periodic.

These periodic variations shift the accumulated discriminator variable $\Phi$ relative to the DC value computed in the prior section. Assuming the filtering operations are linear, as is the case for typical implementations of the noncoherent averaging and loop filter blocks in Fig. 3, then we can equivalently analyze the accumulated discriminator first (prior to filtering) and treat the accumulated discriminator as the input to a subsequent tracking loop.

The periodic deviations of the discriminator output from DC are here labeled $\delta\psi$, where $\delta\psi = (\delta\Phi - \langle\delta\Phi\rangle)$. Evaluating this difference using by subtracting (36) from (33) and substituting (37) gives

$$\delta\psi = \operatorname{atan2}\left(\frac{(B_0^2 - B_1^2)\sin(\Delta\gamma)}{(B_0^2 + B_1^2)\cos(\Delta\gamma) + 2B_0B_1\cos(\delta)}\right) - \Delta\gamma f(\beta, \Delta\gamma) \tag{42}$$

The amplitude ratio $\beta$ factors strongly influences the periodic behavior of the deviation $\delta\psi$ and its accumulation $\psi$. To see this, it is useful to plot the total Doppler estimate (deviation plus DC component) as a periodic function of time, as shown in Fig. 5. The horizontal axis of the figure is normalized time, with time divided by the beat period $T_{beat} = 2\pi/\omega_{beat}$, so that each integer represents a complete period of oscillation. The vertical axis represents the normalized accumulated discriminator output $\Phi/\omega_{av}$. Here the accumulated discriminator $\Phi = \langle\Phi\rangle + \psi$ is computed by integrating the loop contributions of (42) in the manner of integral (8); the normalization is the average of the two arriving frequencies $\omega_{av} = \frac{1}{2}(\omega_{d,0} + \omega_{d,1})$. The signal generally moves only on one side of this average frequency, either above or below, without crossing. The result in Fig. 5 is fully general except for the value of the difference between the two arriving frequencies; this difference has (arbitrarily) been set so that the normalized beat frequency is $\omega_{beat}/\omega_{av} = 1.6\%$. The plot looks similar for other values of beat frequency, but with vertical deviations moving farther from unity as the ratio $\omega_{beat}/\omega_{av}$ becomes larger.

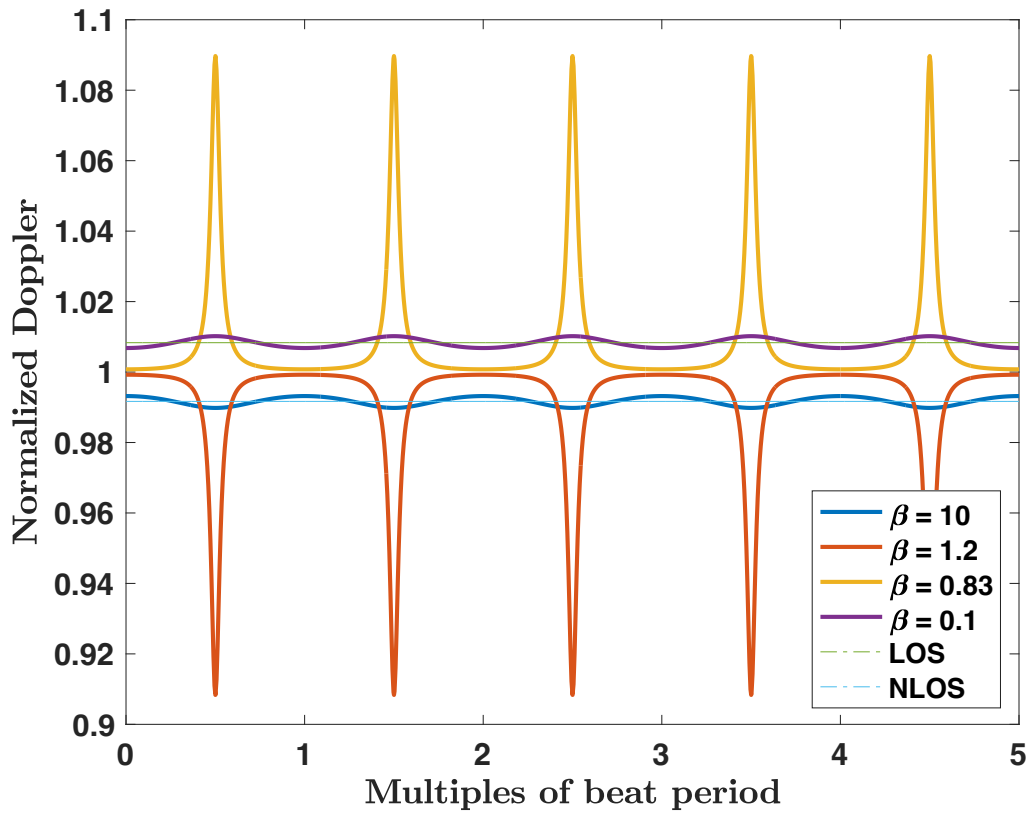

Fig. 5. Normalized discriminator output for a range of relative powers $\beta$, with $\omega_{beat}/\omega_{av} = 1.6\%$.

Let's look closer at the plot over a single beat period $T_{beat} = 2\pi/\omega_{beat}$, as is shown in Fig. 6 and Fig. 7. When the NLOS signal power dominates ($\beta$>1), the estimated Doppler $\tilde{\omega}_d$ is clearly centered on the NLOS signal frequency, as shown on the left side of Fig. 6. The dashed line shows the extreme case ($\beta \to \infty$), where the normalized discriminator output is a constant equal to $\omega_{d,1}$, as is consistent with Remark 1. In this extreme, the signal is dominated by its DC component.

As $\beta$ falls toward one from above, a sinusoid-like ripple appears, which is clearly visible when $\beta = 1.2$. As the power continues to drop, the median value of the deviation shifts upward (toward the normalized average frequency, which is 1, but a sharp downward spike emerges in the middle of the period. The spike results when the denominator of (36) becomes vanishingly small, which happens when $\cos(\delta) = -\cos(\Delta\gamma)$. This happens once per beat period, since $-\cos(\Delta\gamma)$ is constant and since $\cos(\delta)$ covers its full range of values during each beat period. When the denominator of (36) approaches zero, the discriminator output $\langle \delta\Phi \rangle$ diverges.

Interestingly, as the power ratio $\beta$ changes, the change in the area under the spike offsets the change in the area under the flatter sections of the transient, such that there is no net change in the overall integral for a single period; this overall integral corresponds to the period-averaged Doppler $\langle \widetilde{\omega}_d \rangle$. This invariance explains why the period-averaged $\langle \widetilde{\omega}_d \rangle$ remains constant as $\beta$ changes, away from $\beta = 1$ (as shown in Fig. 4).

The special case at $\beta = 1$ is particularly interesting, because a singularity occurs here. In this case, the denominator of (36) goes precisely to zero (at least in the absence of noise), and the spike becomes a delta function. The sign of the delta function reverses suddenly on either side of the singularity, such that the spike direction switches between the case of $\beta \rightarrow 1$ from above and the case of $\beta \rightarrow 1$ from below. The spikes for both cases are shown in Fig. 7. The high and low power ratio cases are, simply, reflections about the average frequency (noted as 1 on the vertical axis); this reflection can be seen by comparing the right and left sides of Fig. 6 and Fig. 7.

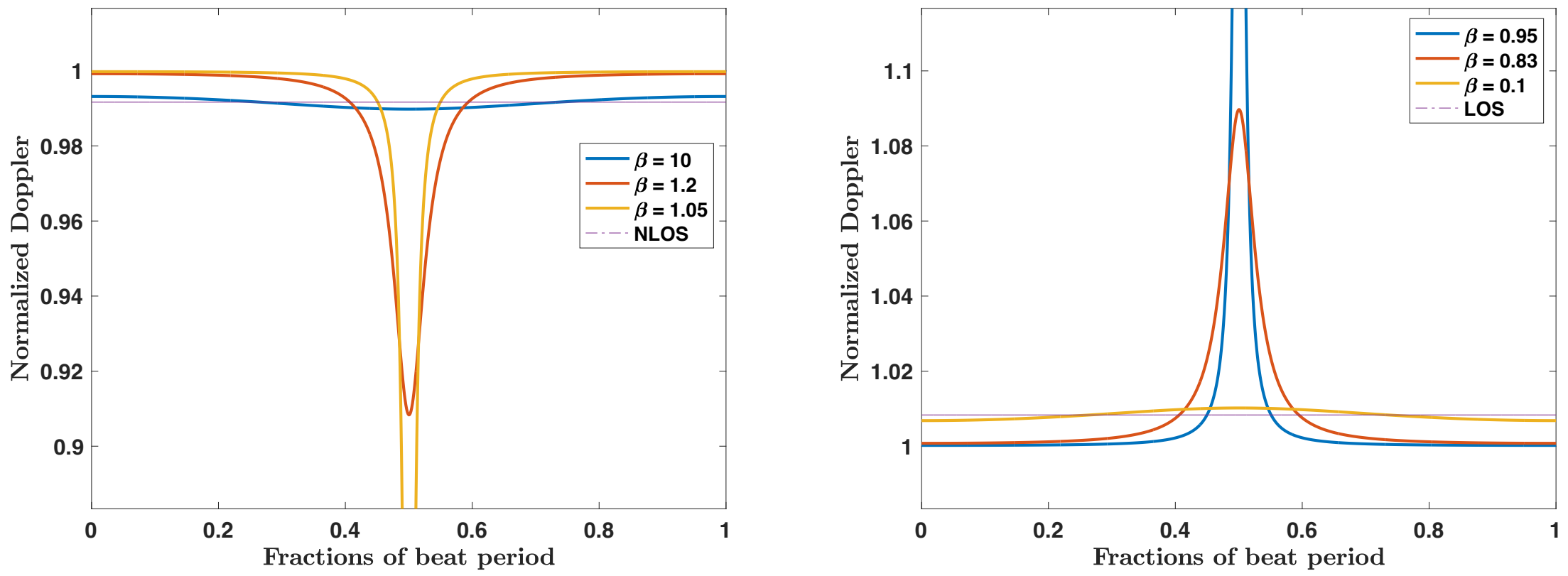

Fig. 6. Estimated Doppler $\widetilde{\omega}_d$ (absent filtering effects) over one period $T_{beat}$ when $\beta > 1$ (left) and $\beta < 1$ (right). The spikes occur at 0.5 period point where the denominator of the arctan function in (41) approaches zero if $\beta$ is close to 1.

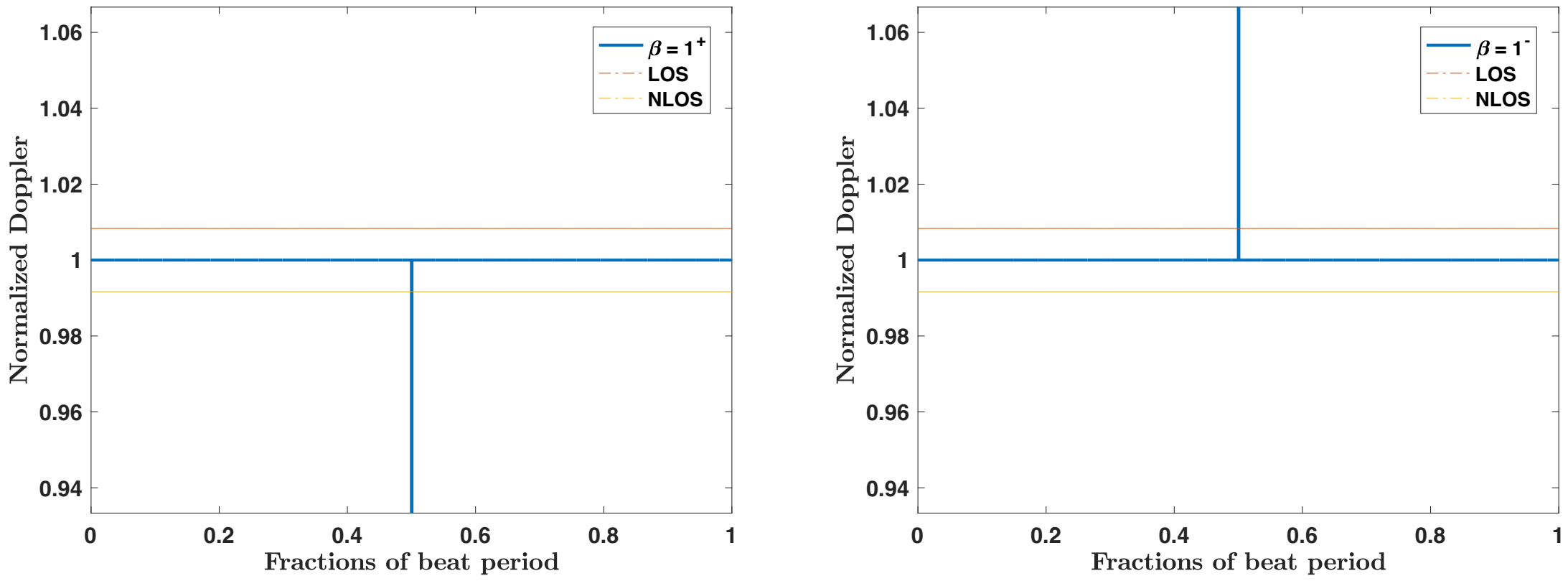

Fig. 7. Estimated Doppler $\widetilde{\omega}_d$ (absent filtering effects) over one period $T_{beat}$ when $\beta \rightarrow 1$ from above (left, noted in legend as $\beta = 1^+$) and from below (right, noted in legend as $\beta = 1^-$). The spikes occur at 0.5 period point where the denominator of the arctan function in (41) goes to zero

As stated previously, the current analysis does not consider noise. When noise effects are considered, the spike observed in the figures becomes less pronounced, and the sharp corners of the step function in Fig. 4 become somewhat smoothed.

From Fig. 6 and Fig. 7, it is clear that the accumulated deviations from DC are distinctly non-sinusoidal, so we must take care to assess their effects. Notably, when filtering effects are considered, the filtering process smooths the sharp spikes that appear in the accumulated deviation $\psi$, decreasing their magnitude and extending their duration. If the filtering process can sufficiently blur the spikes, the FLL-derived discriminator will essentially be a constant value (for constant receiver velocity), corresponding to the DC value quantified in (41). The accumulated deviation $\psi$ will only influence the tracking loop if the loop's bandwidth is somewhat higher than the beat frequency, such that the periodic deviations pass through the filtering process.

Noting that the response of the tracking loop to the accumulated deviation is dependent on the relationship of the beat frequency to filter bandwidth, we will consider an extreme case in our subsequent analysis. Particularly, we will consider the special case when the two rays have equal power *and* when the beat frequency is much lower than the filter bandwidth.

To understand why the proposed extreme is a worst case, first consider the criterion that the two rays have the same power ($\beta \to 1$). When $\beta$ is far from one, the ripples in the discriminator output are small, and meaning that the deviations have negligible amplitude and can be neglected. As an example, consider the left side of Fig. 6, which shows the case for a high amplitude ratio ($\beta = 10$). The ripple is small enough compared to the expected output that the period-averaged result (dashed line) is a very good approximation, regardless of the beat frequency. By contrast, when $\beta$ approaches 1, the amplitude of the perturbation becomes very large. In fact, in the extreme case ($\beta \to 1$), the perturbations will appear to be a chain of Dirac delta functions (impulses), occurring once per each beat period. We will call this series of delta functions a *spike chain*. These spikes will cause the largest possible deviations in the estimated Doppler frequency over the range of possible values of the amplitude ratio $\beta$.

The extreme case also considers the case of a low beat frequency, as compared to filter bandwidth. In the case of a low beat frequency (or long beat period), the filter output responds to each input spike and then decays fully away. As the beat period becomes shorter, the response does not decay as completely, such that the peak-to-trough variations of the output signal are not as large. For simplicity, let us analyze the most extreme case of a low-frequency signal, when the filter output fully decays between spikes.

For this simple model of the extreme case, we need only to know the initial value of the impulse response, noting that the response decays to zero before the next spike appears. To obtain this initial response, we first need to know the magnitude of each spike. By inspection of the step function in Fig. 4, we note that the integral of the deviation is constant for all values of $\beta$ on the same side of the singularity at $\beta = 1$. The integral of the deviation over one period is the area under the curve shown in Fig. 6. The area is equal to the average "height" of the curve multiplied by one beat period $T_{beat} = 2\pi/\omega_{beat}$. The average "height" is the perturbation from the midpoint of the step function in Fig. 4, or $\frac{1}{2}\omega_{beat}$. Thus, the area under the curve (which is preserved as the area is condensed under the spike) is equal to the product $T_{beat} \cdot \frac{1}{2}\omega_{beat} = \pi$ with angular frequency units (rad/sec). The sign of each spike is negative when the dominant frequency is lower than the average frequency and positive when the dominant frequency is higher than the average frequency. For constant-velocity driving, the spike chain is a series of unit delta functions scaled by a magnitude of $\pi$, with $\frac{\Phi}{\Delta t} = \pm \sum_k \delta(t + kT_b)\pi$.

The filter output can be computed as the convolution of the filter impulse function *h*(*t*) with the spike chain. If the filter time constant is shorter than the beat period ($\tau < T_b$), then the output of the loop filter will track the spike chain. In other words, in the $\tau < T_b$ case, the output of the filter will spike sharply and then decay back to a baseline value of $\omega_{av} = \frac{\omega_{d,0}+\omega_{d,1}}{2}$.

To proceed further, let us assume a representative model for the combined filtering that converts the discriminator output into the FLL-derived observable. Specifically, let's model the filtering process as a first-order filter with a bandwidth of 1 Hz. This filter has a time constant of $\tau = \frac{1}{2\pi}$ s. Convolving this filter's impulse function with the spike chain, we obtain the result that each spike increases the filter output by $\Delta\tilde{\omega} = 2\pi^2$ rad/s. We can convert this frequency jump to an equivalent speed jump using (12). This is to say that an impulse-induced jump in frequency of $2\pi^2$ rad/s is equivalent to a jump in estimated projected speed of $\Delta v = 0.60$ m/s. The initial perturbation decays away, but new perturbations are added each beat period. The result is that several of these decayed bumps can accumulate to produce a net perturbation over time, allowing the system to eventually reach its steady state value. Once the system converges to a periodic steady state, the net perturbation from the DC level can be described in terms of minimum and maximum perturbation values. Fig. 8 shows the transient response of the FLL loop filter when suddenly exposed to a chain of delta functions (or *spikes*). The error introduced by the spikes accumulates until the system reaches a periodic steady-state response, which resembles a sawtooth pattern.

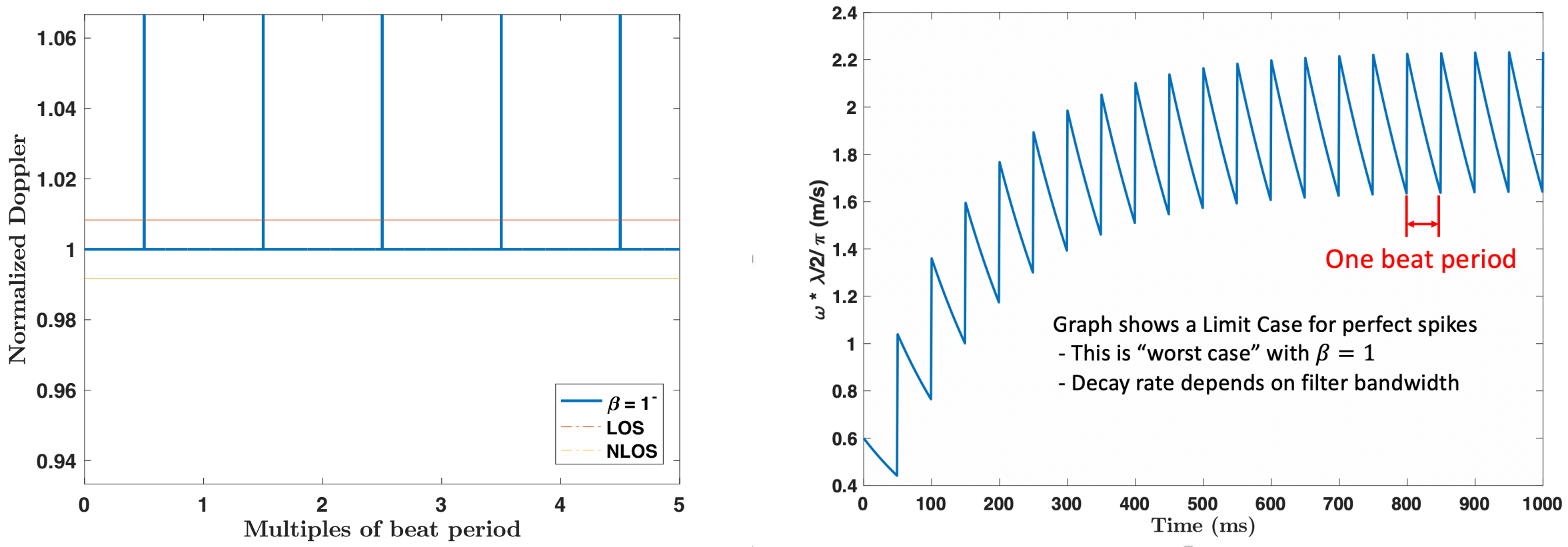

Fig. 8. In the most extreme case, the input oscillations of Fig. 5 can be modeled as a spike chain (left); when the beat period is short compared to the filter time constant, the filter output response (right) resembles a sawtooth pattern superposed on a classic first-order step response

Minimum and maximum perturbations are illustrated in Fig. 9, both in an absolute sense (top) and in a relative sense (bottom). The absolute plot shows perturbations in velocity units (m/s) as a function of beat period $T_b$. In the plot, the top of the shaded region shows the maximum value of the perturbation over each period, and the bottom of the shaded region shows the minimum. Values on the vertical axis are referenced to the period-averaged result, with positive perturbations shifted toward the average Doppler and negative perturbations shifted away from the averaged Doppler. The difference between the top and bottom is always equal to the step change injected by a single impulse ($\Delta v = 0.60$ m/s), regardless of beat period. The deviation from the period-averaged reference does shift with beat period, however, progressing from symmetric variations about the reference (on the left side of the plot, where the beat period is short and smoothing is effective) to one-sided variations (on the right side of the plot, where the beat period is long and smoothing is ineffective).

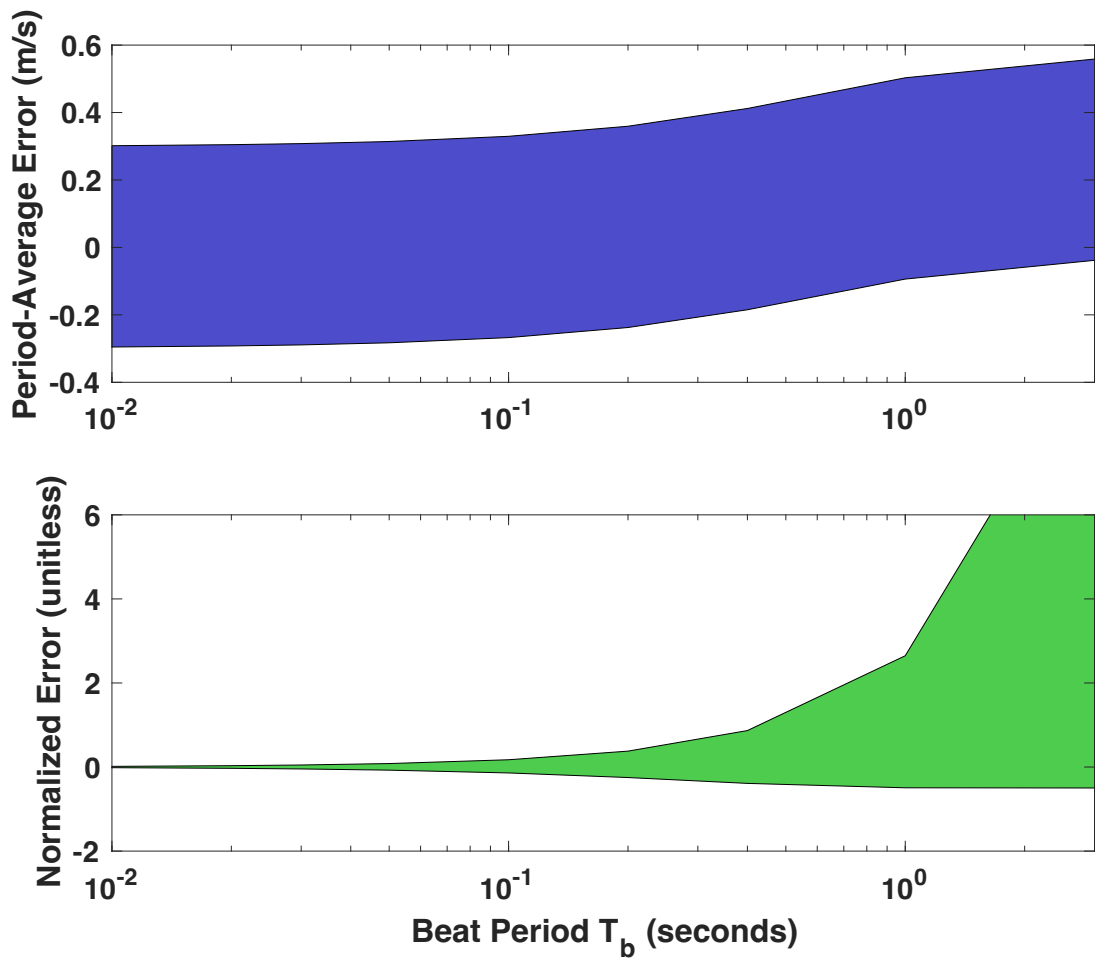

Fig. 9. Bounds on period deviations for loop filter in absolute terms (m/s, top) and in normalized terms (unitless, bottom)

Although the absolute magnitude of the perturbations is always the same, as shown on the top of Fig. 9, the relative magnitude of these perturbations as compared to the difference between LOS and NLOS outputs is a strong function of the beat period. To study the relative effects, we normalize the perturbation in the FLL-observable by the speed difference between the LOS-only and NLOS results. Normalized perturbations are plotted as a function of beat period $T_b$ on the bottom of Fig. 9. The normalization term (difference between outcomes for LOS and NLOS cases) is inversely proportional to $T_b$. As such, the normalization is very large when $T_b$ is small, which means that the perturbations are very small compared to the NLOS errors in the short $T_b$ case (when smoothing is effective). The size of the perturbations is very large compared to the normalization, however, for large $T_b$ (when smoothing is not effective).

**DISCUSSION**

This paper focuses on a theoretical characterization of multipath effects on FLLs, the paper results are also supported by experimental studies. In particular, the three remarks enumerated above are experimentally verified.

The first remark (NLOS case) suggests that the projected velocity estimate obtained from the FLL Doppler observable will be correct within a scale factor. This can be seen by using (13) to construct the ratio of the NLOS observable to the LOS-only observable:

$$\frac{\omega_{r,NLOS}}{\omega_{r,LOS}} = \frac{\cos(\theta_{r,NLOS})}{\cos(\theta_{r,LOS})} \tag{43}$$

Recall that $\theta$ refers to the angle between the direction of propagation of the signal and the direction of motion of the receiver. In general, the incoming directions are different for the NLOS and LOS signal components (that is, $\theta_{r,NLOS} \neq \theta_{r,LOS}$), and so the NLOS observable is a scaled version of the LOS observable. The first remark is easily testable in an experiment if the actual receiver motion is known and the receiver speed is slowly varied. In the case of an NLOS signal, the projected Doppler velocity should have the same shape but not the same magnitude as the ground truth. Also, an important implication is that NLOS signals may potentially be identified by using the ratio (43), as discussed in [15].

The second remark (zero-speed multipath case) suggests that measurements should not be biased when the receiver is stationary. This remark is easily testable in an experiment if the receiver speed is zero. If multipath is present, random noise is expected, but the mean of the projected-speed distribution should be zero.

The third remark (nonzero-speed multipath case) suggests that the Doppler observable in a high-multipath environment should either be correct, matching the expected observable for the LOS-only case, or be scaled according to (43), where the scale factor is associated with the incident direction of a strong NLOS signal. This remark is potentially testable by creating an experiment in which an NLOS signal is present and in which receiver motion causes the power of the LOS signal to drop from a high level to a very low level (for example, by moving the receiver behind a building which occludes the LOS signal but not the NLOS signal). In such an experiment, we would expect the Doppler observable to transition rapidly from tracking the LOS signal to the NLOS signal predicted by the step function shown in Fig. 4. Strictly speaking, we expect this sudden switch would be easily observable only if the beat frequency were high enough to be attenuated by the loop filter. By contrast, we would expect the instantaneous Doppler estimate to oscillate continuously if the beat frequency were lower than the bandwidth of the FLL tracking loop. Such low frequency beats may occur when the incidence angle is very small, or the vehicle speed is very low. This can be seen by substituting (13) into (34) to obtain

$$\omega_{beat} = |\omega_{d,0} - \omega_{d,1}| = \frac{2\pi\|\mathbf{v}_r\|}{\lambda}\left|\cos(\theta_{r,0}) - \cos(\theta_{r,1})\right| \tag{44}$$

This equation clearly links the beat frequency to physical factors, specifically receiver velocity and the incident raypath angles of the LOS and NLOS signals.

A potentially interesting application of this modeling effort is to suggest that multipath signals might be used to create new signals of opportunity for enhanced velocity estimation. There are existing applications of multipath signals in GNSS reflectometry [25, 26]. Recent work has also explored multipath as a signal of opportunity in urban positioning, via shadow matching or pseudorange reconstruction with assistance of 3D map [27, 28]. However, the utilization of Doppler multipath as signals of opportunity in velocity estimation has not yet been investigated. Using multipath signals in this way requires a specialized antenna that separates out the LOS and NLOS components and that infers the angle of arrival for each [16]. For instance, a MIMO antenna could be used to create multiple beams focused separately on the LOS and NLOS signals. The angles of those beams relative to the receiver could be inferred from the beamforming algorithm, and those angles could in turn be mapped into the inertial frame if IMU data were also available. The result would be a system capable of evaluating (13) separately for each ray path, implying that the NLOS Doppler components would provide independent information to supplement the LOS Doppler component for each navigation beacon (satellite or ground-based). A preliminary algorithm for processing such signals of opportunity is described in [10].

Many aspects of this work might be expanded in the future. Future work topics of particular significance are listed below.

1) Multipath effects on the FLL have not yet been investigated when more than two rays arrive at the user receiver. It is natural to extend the model in this dissertation to the case where the multipath signal involves more than two signal components, to verify the generalizability of the results in this paper. Also, future work should further investigate the role of random noise during FLL tracking, accounting for the nonlinear effects of the discriminator.

2) Different FLL design options have not yet been investigated (e.g. discriminators other than the arctangent discriminator). Though we believe this design analyzed in this paper is representative of a wide range of FLL designs on the market, it is worth verifying this hypothesis.

3) Applications of the model should be explored. The model described in this paper can potentially be applied in antispoofing applications for GPS [29, 30, 31, 32], for example, where the authentic and the spoofing signal are two incoming signal sources. In concept, inferred angles-of-arrival might be employed to distinguish spoofing signals that are strong relative to the LOS signal.

4) Verify these concepts with controlled experiments. Preliminary experimental results are presented, but more comprehensive experimental data are needed to gain insight into urban multipath events, in order to ascertain the relative power levels of LOS and NLOS signals. A simulator is also needed to generate a perfectly controlled two-ray multipath signal for further testing on Remark 3.

## CONCLUSION

This paper investigates the impact of NLOS and multipath signals on carrier-frequency tracking by the FLL. NLOS and multipath effects are analyzed theoretically, to model their effects on the Doppler observable. Our analysis leads to three key observations, based on an assumption that NLOS signals are reflected from nearby, stationary surfaces. First, in the case when only one NLOS signal component is tracked, the Doppler observable will be proportional to receiver speed, but the scale factor will be wrong. Second, in the case when a stationary receiver experiences multipath, the expected value of the Doppler observable will be correct, indicating the receiver speed is zero. Third, when a moving receiver experiences multipath, the FLL will tend to track the strongest received signal component (LOS or NLOS) as if the other signal components are absent. The last case has a caveat, which is to say that the moving receiver's FLL may experience additional time variations due to multipath, with those variations being periodic at a predictable beat frequency when the receiver speed is constant. The three analytically-based observations were used to interpret experimentally acquired data. Though the details of the receiver hardware design were not fully known, the three analytically-based observations provided a qualitative interpretation of the data.

## APPENDIX 1 Derivation from precorrelation signal to postcorrelation signal

This Appendix clarifies the formulation of the postcorrelation signal $S(t)$ in (4) from the precorrelation signal $s(t)$ in (3). In particular, we focus on an analysis for the two-ray case without noise. For this case, the signal model (3) can be written with only two terms:

$$s(t) = A_0 e^{j\theta_0} c(t - \tau_0) + A_1 e^{j\theta_1} c(t - \tau_1) \tag{45}$$

The local carrier and code replica $s_R$ is given by

$$s_R(t) = e^{j\theta} c(t - \tau) \tag{46}$$

Here the total, time-varying phase is represented as $\theta$ noting that $\theta = (\omega + \tilde{\omega}_d)t + \tilde{\phi}$. The mixing process gives

$$ss_R' = A_0 e^{j(\theta_0 - \theta)} c(t - \tau_0) c(t - \tau) + A_1 e^{j(\theta_1 - \theta)} c(t - \tau_1) c(t - \tau) \tag{47}$$

The exponential terms here are sinusoids that oscillate at a low frequency (e.g. at the frequency of the Doppler error $\delta\omega_l$ between the replica and the $l$-th component of the received signal).

$$e^{j(\theta_l - \theta)} = e^{j(\delta\omega_l)t + \phi_l} \tag{48}$$

Here the term $\phi_l$ is the unknown phase shift associated with the signal; this phase shift is not tracked by an FLL (though it would be tracked by a PLL). Over a single integration interval $T$ (of about 1 ms), the exponential value in (48) is approximately constant as long as the Doppler estimation period $(2\pi/\delta\omega)$ is long compared to the integration interval $T$. This approximation is reasonable for our application, where the Doppler estimation error is dominated by the beat oscillation, which is attenuated by post-discriminator filtering unless the beat period is very long (much longer than 10 $ms$ according to Fig. 9).

Integration creates the postcorrelation signal S, described by the following integral

$$S(t) = \frac{1}{T}\int_0^T ss_R' \, dt = \frac{1}{T}\int_0^T \left(A_0 e^{j(\theta_0-\theta)} c(t-\tau_0)c(t-\tau) + A_1 e^{j(\theta_1-\theta)} c(t-\tau_1)c(t-\tau)\right) dt \tag{49}$$

Since the exponential term (48) can be modeled as approximately constant for our application, the integral (49) can be evaluated as

$$S(t) = A_0 e^{j((\delta\omega_0)t+\phi_0)} R_0(\tau) + A_1 e^{j((\delta\omega_1)t+\phi_1)} R_1(\tau) \tag{50}$$

where

$$R_l(\tau) = \frac{1}{T}\int_0^T c(t-\tau_l)c(t-\tau) \, dt, \;\; l = 0,1 \tag{51}$$

To keep our model of the postcorrelation signal concise, we can rewrite (50) as

$$S(t) = B_0 e^{j((\delta\omega_0)t+\phi_0)} + B_1 e^{j((\delta\omega_1)t+\phi_1)} \tag{52}$$

Here the amplitude values, which are relabeled $B_l$, are the products of the input amplitudes $A_l$ and the correlation values $R_l$.

Though we have derived this model for the two-ray case, it generalizes trivially to the multi-ray case. This generalization is listed in the main body of this article as equation (4).

**APPENDIX 2 Discriminator algorithms in complex numbers and in real numbers**

Different from the discriminator algorithms defined in real numbers in the traditional textbook [6], in this paper, the discriminator algorithm is expressed in complex number form for the convenience of analysis. These two forms are mathematically equivalent. For comparison, if the complex postcorrelation signal is written as $S = I + Qi$, where $i^2 = -1$, then Table 1 below puts two expressions together for the reference.

Table 1. Discriminator algorithms in real numbers and their equivalences in complex numbers

| | FLL | PLL |
|---|---|---|
| Discriminator algorithm in real numbers | $\frac{atan2(cross, dot)}{\Delta t}$ where $dot = I_{k-1} \times I_k + Q_{k-1} \times Q_k$, $cross = I_{k-1} \times Q_k - I_k \times Q_{k-1}$, $\Delta t = t_k - t_{k-1}$ | $atan(\frac{Q_k}{I_k})$ |
| Discriminator algorithm in complex numbers | $\frac{\angle S(t)S'(t-\Delta t)}{\Delta t}$ | $\angle S(t)$ |